\documentclass[aps,prl,twocolumn,groupedaddress,floats,showpacs]{revtex4}
\usepackage{latexsym}
\usepackage{dcolumn}
\usepackage[dvips]{graphicx}
\usepackage{amssymb}
\usepackage{graphics}
\usepackage{amsmath}
\usepackage{epsf}

\begin{document}

\author{C. Jang$^{1}$, S. Adam$^{2}$, J.-H. Chen$^{1,3}$,
E. D. Williams$^{1,3}$, S. Das Sarma$^{1,2}$ and M. S. Fuhrer$^{1,3}$}

\affiliation{$^1$Center for Nanophysics and Advanced Materials,}

\affiliation{$^2$Condensed Matter Theory Center,}

\affiliation{$^3$ Materials Research Science and Engineering Center,
Department of Physics, University of Maryland, College Park, MD
20742-4111, USA}

\title{Tuning the effective fine structure constant
in graphene: opposing effects of dielectric screening on short- and
long-range potential scattering}
        
\date{\today}
\begin{abstract}
We reduce the dimensionless interaction strength $\alpha$ in
graphene by adding a water overlayer in ultra-high vacuum,
thereby increasing dielectric screening.  The
mobility limited by long-range impurity scattering is {\it increased} over 30
percent, due to the background dielectric constant enhancement
leading to reduced interaction of electrons with charged
impurities.  However, the carrier-density-independent conductivity due
to short range impurities is {\it decreased} by almost 40 percent, due to
reduced screening of the impurity potential by conduction electrons.
The minimum conductivity is nearly unchanged, due to canceling
contributions from the electron/hole puddle density and long-range
impurity mobility.  Experimental data are compared with theoretical
predictions with excellent agreement.
\end{abstract}
\pacs{}
\maketitle

Most theoretical and experimental work on graphene has focused on its
gapless, linear electronic energy dispersion $E= \hbar v_{\rm F} k$.
One important consequence of this linear spectrum is that the
dimensionless coupling constant $\alpha$ (or equivalently $r_s$,
defined here as the ratio between the graphene Coulomb potential
energy and kinetic energy) is a carrier-density independent
constant~\cite{kn:peres2005,kn:ando2006}, and as a
result, the Coulomb potential of charged impurities in graphene is
renormalized by screening, but strictly maintains its long-range
character.  Thus there is a clear dichotomy between long-range and
short-range scattering in graphene, with the former giving rise to a
conductivity linear~\cite{kn:ando2006,kn:hwang2006c} in
carrier density (constant mobility), and the latter having a constant
conductivity independent of carrier density.  Charged impurity
scattering necessarily dominates at low carrier density, and the
minimum conductivity at charge neutrality is determined by the charged
impurity scattering and the self-consistent electron and hole puddles
of the screened impurity
potential~\cite{kn:hwang2006c,kn:adam2007a,kn:tan2007,kn:chen2008}.

Apart from the linear spectrum, an additional striking aspect of
graphene, setting it apart from all other two-dimensional electron
systems, is that the electrons are confined to a plane of atomic
thickness.  This fact has a number of ramifications which are only
beginning to be explored~\cite{kn:min2008}.  One such
consequence is that graphene's properties may be tuned enormously by
changing the surrounding environment.  Here we
provide a clear demonstration of this by reducing the dimensionless
coupling constant $\alpha$ in graphene by more than 30 percent through
the addition of a dielectric layer (ice) on top of the graphene sheet.
Upon addition of the ice layer, the mobility limited by long-range
scattering by charged impurities {\it increases} by 31 percent, while
the conductivity limited by short-range scatterers {\it decreases} by
38 percent.  The minimum conductivity value remains nearly unchanged.
The opposing effects of reducing $\alpha$ on short-and long-range
scattering are easily understood theoretically.  The major effect on
long-range scattering is to reduce the Coulomb interaction of
electrons with charged impurities, reducing the scattering
~\cite{kn:jena2007}.  In
contrast, the dielectric does not modify the atomic-scale potential of
short-range scatterers, and there the leading effect is the reduction
of screening by the charge carriers, which increases scattering
resulting in lower high-density conductivity.  Such screening of 
short-range potentials has been
predicted theoretically~\cite{kn:ando1982}, although
in other $2D$ systems, this effect is difficult to observe
experimentally.  The minimum conductivity is nearly unchanged due to
competing effects of increased mobility and reduced carrier
concentration in electron-hole puddles due to reduced
screening~\cite{kn:adam2007a,kn:rossi2008}.

\begin{figure}
\bigskip
\epsfxsize=0.8\hsize
\hspace{0.0\hsize}
\epsffile{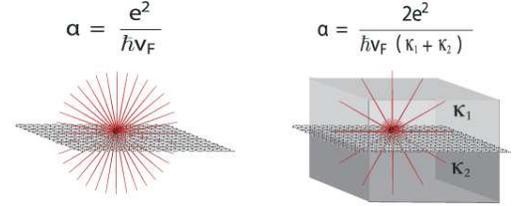}
\caption{\label{Fig:cartoon}  Schematic illustrating dielectric
screening in graphene.  The dielectric environment controls 
in the interaction strength parameterized by the coupling constant $\alpha$.}
\end{figure}

Fig.~\ref{Fig:cartoon} illustrates the effect of the dielectric
environment on graphene.  For graphene sandwiched between two
dielectric slabs with $\kappa_1$ and $\kappa_2$,
\begin{equation}
\alpha = \frac{2 e^2}{(\kappa_1 + \kappa_2) \hbar v_{\rm F}},
\end{equation}
where $e$ is the electronic charge, $\hbar$ is Planck's constant, and
$v_{\rm F}$ is the Fermi velocity, which we take to be
$1.1\times 
10^{6}~\mbox{\rm m/s}$~\cite{kn:novoselov2005,kn:zhang2005,kn:jiang2007}.
Typically, graphene transport 
experiments~\cite{kn:novoselov2005,kn:zhang2005,kn:tan2007,kn:chen2008}
are performed on a SiO$_2$ substrate with $\kappa_1 \approx 3.9$ and
in air/vacuum $\kappa_2 \approx 1$, making graphene a weakly
interacting electron system with $\alpha \approx 0.8$ (although very
recently work on substrate-free graphene~\cite{kn:bolotin2008}
explored the strong coupling regime with $\alpha \approx 2$).
Here we deposit ice ($\kappa_2 \approx 3.2$~\cite{kn:petrenko1999}) on
graphene on SiO$_2$, decreasing $\alpha$ from $\approx 0.81$ to
$\approx 0.56$.

\begin{figure}
\bigskip
\epsfxsize=0.75\hsize
\hspace{0.0\hsize}
\epsffile{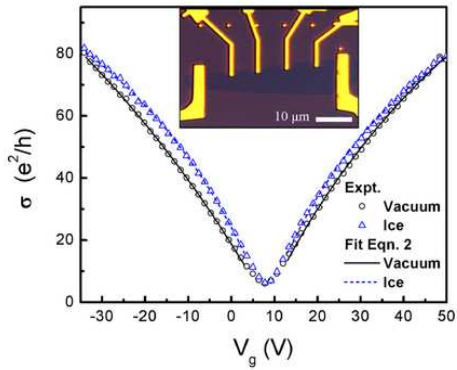}
\caption{\label{Fig:Conductivity} (Color online) Conductivity 
of the graphene device as a function of back-gate voltage 
for pristine graphene (circles) and after deposition 
of 6 monolayers of ice (triangles).  Lines are
fits to Eq.~\ref{Eq:MattRule}. Inset: Optical microscope image of the device.}
\end{figure}

\begin{figure}
\bigskip
\epsfxsize=0.7\hsize
\hspace{0.0\hsize}
\epsffile{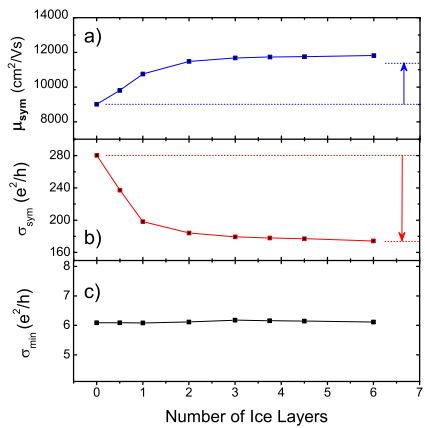}
\caption{\label{Fig:Compare} (Color online) $\mu_{\rm sym}$,
$\sigma_{\rm sym}$ and $\sigma_{\rm min}$ as a function of number of
ice layers.  Dashed lines show the values for pristine graphene
and corresponding theoretical expectations for the ice-covered device.}
\end{figure}

Graphene is obtained by mechanical exfoliation of Kish graphite on a
SiO$_2$ (300 {\rm nm})/Si substrate~\cite{kn:novoselov2005}.  
Graphene
monolayers are identified from the color contrast in an optical
microscope image and confirmed by Raman 
spectroscopy~\cite{kn:ferrari2006}.  The final device (see
Fig~\ref{Fig:Conductivity} inset) was fabricated by patterning
electrodes using electron beam lithography and thermally evaporated
Cr/Au, followed by annealing in Ar/H$_2$ to remove resist residue (see
Refs.~\cite{kn:ishigami2007,kn:chen2008} for details).  The experiments are performed
in a cryostat cold finger placed in an ultra high vacuum (UHV)
chamber.  In order to remove residual adsorbed gases on the device and
the substrate, the sample was baked at $430~{\rm K}$ over-night in UHV
following a vacuum bakeout.  The conductivity was measured using a
conventional four-probe technique with an ac current of $50~{\rm nA}$
at a base pressure ($\sim 10^{-10}~{\rm torr}$) and device temperature
($\sim 77~{\rm K}$).  Deionized nano-pure water was introduced through
a leak valve attached to the chamber.  The water gas pressure
(determined by a residual gas analyzer) was $5 \pm 3 \times
10^{-8}~{\rm torr}$.  The amount of ice deposited was estimated
by assuming a sticking coefficient of unity and the ice I$_h$ layer
density of 
$9.54 \times 10^{14}~{\rm cm}^{-2}$~\cite{kn:sanfelix2003,kn:thiel1987}.
 
Fig.~\ref{Fig:Conductivity} shows conductivity as a function of gate
voltage for two different sample conditions, pristine graphene and
ice-covered graphene.  We observe several interesting effects of
adding ice: (i) The offset gate voltage at which the conductivity is a
minimum $V_{g, {\rm min}}$ remains unchanged; (ii) the minimum
conductivity $\sigma_{\rm min}$ value remains unchanged, (iii) the
maximum slope of $\sigma(V_{g})$ becomes steeper, and (iv) the curve
$\sigma(V_{g})$ in the presence of ice is more non-linear and crosses
that of the pristine sample at some large carrier density.  All these
features can be understood qualitatively from the physical picture
described above, and we show below that they are in quantitative
agreement with the predictions of Boltzmann transport theory including
screening within the Random Phase Approximation (RPA).

In order to interpret the experimental results
quantitatively~\cite{kn:morozov2008}, we fit the conductivity data to

\begin{equation}
\sigma^{-1}(V_g, \alpha) = (n e \mu_l)^{-1} + (\sigma_s)^{-1},
\label{Eq:MattRule}
\end{equation}
where $n = c_g |V_g - V_{g, {\rm min}}|$, $e$ is the electric charge
and $c_g = 1.15 \times 10^{-8}~{\rm V/cm^2}$ is the gate capacitance
per unit area for the $300~{\rm nm}$ thick SiO$_2$.  Since the
transport curves are not symmetric about the minimum gate voltage, the
fitting is performed separately for positive and negative carrier
densities (i.e. electron and hole carriers), excluding data close to
the Dirac point conductivity plateau $(V_{g, {\rm min}} \pm 5V)$.  We
report both the symmetric $\mu_{\rm sym} (\sigma_{\rm sym})$ and
antisymmetric $\mu_{\rm asym} (\sigma_{\rm asym})$ contributions to
the mobility (conductivity).  Shown also in Fig.~\ref{Fig:Conductivity} 
is the result of the fit for pristine graphene and after deposition 
of $6$ monolayers of ice.

Figure~\ref{Fig:Compare} shows $\mu_{\rm sym}$,
$\sigma_{\rm sym}$ and $\sigma_{\rm min}$ as a function of number of ice layers.  
The mobility (Fig.~\ref{Fig:Compare}a) of pristine
graphene is $9,000~{\rm cm^2/Vs}$,
which is typical for clean graphene devices on SiO$_2$ substrates at
low temperature.  As the number of water layers increases, 
the mobility {\it increases}, and saturates after about $3$ layers
of ice to about $12,000~{\rm cm^2/Vs}$.  In 
contrast, the conductivity due to short-range scatterers 
(Fig.~\ref{Fig:Compare}b) decreases from $280~e^2/h$ to $170~e^2/h$.
The decrease in conductivity due to short-range scatterers shows
a similar saturation behavior as the mobility, suggesting they
have the same origin~\cite{kn:footnote1}.  The absence of any sharp change in 
the conductivity or mobility at very low ice coverage rules out
ice itself acting as a significant source of short- or long-range
scattering.  This is corroborated by the absence of a shift in the gate 
voltage of the minimum conductivity, consistent with 
physisorbed ice~\cite{kn:sanfelix2003} not donating charge to 
graphene~\cite{kn:adam2007a,kn:tan2007,kn:chen2008}.
Fig.~\ref{Fig:Compare}c shows that the minimum conductivity is
essentially unchanged during the addition of ice.

\begin{widetext}
\vspace{-0.3in}
\begin{center}
\begin{table}[h!]
\caption{\label{Tab:Table} Summary of our results 
and corresponding theoretical predictions.}
\vspace{0.2in}
\begin{tabular}{||c|c|c|c||}
\hline \hline
& \hspace{0.55in} & \hspace{0.55in}  & \hspace{0.55in}   \\ 
\hspace{0.55in} &  \hspace{0.55in} 
& \hspace{0.1in} Theory \hspace{0.1in} & 
\hspace{0.1in} Experiment \hspace{0.1in} \\ 
& & & \\
\hline 
Long-range (symmetric): 
$\frac{\mu_{\rm sym}^{\rm ice}}{\mu_{\rm sym}^{\rm vac}}
= \frac{F_l(\alpha^{\rm vac})}{F_l(\alpha^{\rm ice})}$
& 
Ref.~\protect{\cite{kn:adam2007a}} & 
$1.26$ &
$1.31$ \\ 
\hline 
Short-range (symmetric): 
$\frac{\sigma_{\rm sym}^{\rm ice}}{\sigma_{\rm sym}^{\rm vac}}
= \frac{F_s(\alpha^{\rm vac})}{F_s(\alpha^{\rm ice})}$ 
&
Ref.~\protect{\cite{kn:adam2007b}} &
$0.62$ &  
$0.62$ \\
\hline
Minimum conductivity: 
$\frac{\sigma_{\rm min}^{\rm ice}}{\sigma_{\rm min}^{\rm vac}}
= \frac{n^*(\alpha^{\rm ice}) F_l(\alpha^{\rm vac})}
{n^*(\alpha^{\rm vac}) F_l(\alpha^{\rm ice})}$  &
Ref.~\protect{\cite{kn:adam2007a}} &
$0.99$ &  
$1.00$ \\
\hline
Long-range (anti-symmetric): 
$\frac{\mu_{\rm asym}^{\rm ice}}{\mu_{\rm asym}^{\rm vac}}
= \frac{F_l(\alpha^{\rm vac})~\alpha^{\rm ice}}{F_l(\alpha^{\rm ice})~\alpha^{\rm vac}}$ 
&
Ref.~\protect{\cite{kn:novikov2007}} &
$0.87$ &
$0.17$ \\
\hline
Short-range (anti-symmetric): 
$\frac{\sigma_{\rm asym}^{\rm ice}}{\sigma_{\rm asym}^{\rm vac}}$ &
Ref.~\cite{kn:huard2008} & &
$0.13$ \\ 
\hline \hline
\end{tabular}
\end{table}
\end{center}
\vspace{-0.3in}
\end{widetext}

We now analyze the experimental results within Boltzmann transport
theory.  The conductivity of graphene depends strongly on the coupling
constant $\alpha$.  For screened long-range impurities within RPA, we
have~\cite{kn:adam2007a}

\begin{eqnarray} 
\sigma_l &=& \frac{2 e^2}{h} \frac{n}{n_{\rm imp}} \frac{1}{F_l(\alpha)},
\nonumber \\
F_l (\alpha) &=& \pi \alpha^2 + 24 \alpha^3 (1 - \pi \alpha) 
                 \nonumber \\
&& \mbox{} 
+ \frac{16 \alpha^3(6 \alpha^2 -1)\arccos[1/2 \alpha]}{\sqrt{4 \alpha^2 -1}},
\label{Eq:Coulomb}
\end{eqnarray}
where in the last term, for $\alpha < 0.5$ both  
$\arccos[(2 \alpha)^{-1}]$ in the numerator and 
$\sqrt{4 \alpha^2 - 1}$ in the denominator are purely 
imaginary so that $F_l(\alpha)$ is real and positive
for all $\alpha$.  For screened short-range impurities, 
we have~\cite{kn:adam2007b}
\begin{eqnarray}
\sigma_s &=& \frac{\sigma_0}{F_s (\alpha)}, \nonumber \\
F_s (\alpha) &=& \frac{\pi}{2} - \frac{32 \alpha}{3} + 
24 \pi \alpha^2 + 320 \alpha^3 (1 - \pi \alpha) \nonumber \\
&& \mbox{}  + 256 \alpha^3 (5 \alpha^2 - 1) 
\frac{\arccos[1/2 \alpha]}{\sqrt{4 \alpha^2 -1}},
\label{Eq:ShortRange}
\end{eqnarray}
where similarly $F_s(\alpha)$ is real and positive.
Consistent with the physical picture outlined earlier, in the limit
$\alpha \rightarrow 0$, $\sigma_l \sim \alpha^{-2}$ which describes
the scaling of the Coulomb scattering matrix element, while for
short-range scattering, $\sigma_s \approx {\rm const} [1 + (64/3 \pi)
\alpha]$ where increased screening of the potential by the carriers
gives the leading order increase in conductivity.  For the experimental
values of $\alpha$, the full functional form of $F_s$ and
$F_l$ should be used~\cite{kn:footnote2}.  Dashed lines in
Figs.~\ref{Fig:Compare}a-b show the theoretical expectations for $\mu_{\rm
sym}$ and $\sigma_{\rm sym}$ for vacuum and ice on
graphene in quantitative agreement with experiment.  

Regarding the magnitude of the minimum conductivity, it was recently
proposed~\cite{kn:adam2007a} that 
one can estimate $\sigma_{\rm min}$ by computing the
Boltzmann conductivity of the residual density $n^*$ that is induced
by the charged impurities.  This residual density 
(i.e. rms density of electrons and hole puddles) has been seen directly in scanning probe
experiments~\cite{kn:martin2008} and in numerical
simulations~\cite{kn:rossi2008}.  We therefore use Eq. 3, but replace
$n$ with $n^*$ = $\langle V_D^2 \rangle /[\pi (\hbar v_{\rm F})^2]$
(where the angular brackets indicate ensemble averaging over
configurations of the disorder potential $V_D$) to
give~\cite{kn:adam2007a}

\begin{eqnarray}
\sigma_{\rm min}  &=&  \frac{2e^2}{h} 
\frac{1}{F_l(\alpha)} \frac{n^*(\alpha)}{n_{\rm imp}}, \nonumber \\
\langle V_D^2 \rangle &=& n_{\rm imp} (\hbar v_{\rm F} \alpha)^2
\int d {\bf q} \left(\frac{e^{-qd}}{q \epsilon(q)}\right)^2, \end{eqnarray}
where $\epsilon(q)$ is the RPA dielectric function and $d \approx 1
{\rm nm}$ is the typical impurity separation from the graphene 
sheet.  The dominant contribution to both the disorder
potential $\langle V_D^2 \rangle$ and $F_l(\alpha)$ is the Coulomb
matrix element, giving $n^* \sim n_{\rm imp} \alpha^2$ and
$1/F_l(\alpha) \sim 1/\alpha^2$ so that to leading order, $\sigma_{\rm min}$
is unchanged by dielectric screening~\cite{kn:footnote3}.

The experimental data also show a mobility asymmetry (between
electrons and holes) of about $10$ percent.
Novikov~\cite{kn:novikov2007} argued that for Coulomb impurities in
graphene such an asymmetry is expected since electrons are slightly
repelled by the negative impurity centers compared to holes resulting
in slightly higher mobility for electrons (since $V_{g, {\rm min}} >
0$, we determine that there are more negatively charged impurity
centers, see also Ref.~\cite{kn:chen2008}); and that for unscreened
Coulomb impurities $\mu_{\rm usc}(\pm V_g) \sim 
[C_2 \alpha^2  \pm C_3 \alpha^3 + C_4\alpha^4 + \cdots]^{-1}$. 
From the magnitude of the asymmetry, we know that 
$C_3 \alpha^3 \ll C_2 \alpha^2$, but if we further 
assume that $C_4 \alpha^4 \ll C_3 \alpha^3$ 
(although, in the current experiment, we cannot extract the value of $C_4$), 
then including the effects of screening gives 
$\mu_{\rm asym} \sim \alpha/F_l(\alpha)$.

In Table~\ref{Tab:Table} we show all the experimental fit parameters
and compare them to theoretical predictions.  The quantitative
agreement for $\mu_{\rm sym}$, $\sigma_{\rm min}$ and $\sigma_{\rm
sym}$ is already highlighted in Fig.~\ref{Fig:Compare}, while we have
only qualitative agreement for $\mu_{\rm asym}$, probably because the
condition $C_4 \alpha^4 \ll C_3 \alpha^3$ does not hold in our
experiments.  There is no theoretical expectation of asymmetry in
$\sigma_s$; the experimental asymmetry (about $30$ percent) could be
explained by contact resistance~\cite{kn:huard2008} which we estimate
to be a 20 percent correction to $\sigma_s$ for our sample geometry.

In conclusion we have observed the effect of dielectric environment on
the transport properties of graphene.  The experiment highlights the
difference between long-range and short-range potential scattering in
graphene.  The enhanced $\mu_l$ (i.e. the slope of $\sigma$ against
density) and reduced $\sigma_s$ (i.e. the constant conductivity 
at high density) are attributed
to the decreased interaction between charged carriers and impurities
and decreased screening by charge carriers, respectively, upon an
increase in background dielectric constant with ice deposition in UHV.
These variations quantitatively agree with theoretical expectations
for the dependence of electron scattering on graphene's ``fine
structure constant'' within the RPA approximation. This detailed
knowledge of the scattering mechanisms in graphene is essential for
design of any useful graphene device, for example, use of a 
high-$\kappa$  gate dielectric will increase the transconductance of graphene at the
expense of linearity, an important consideration for analog
applications.  As demonstrated here, dielectric deposition only
improved mobility by $30$ percent, however the use of high-k
dielectric overlayers could significantly enhance this result.

We thank E. Hwang and E. Rossi for fruitful discussions.  This work is
supported by US ONR, NRI-SWAN and NSF-UMD-MRSEC grant DMR 05-20471.


\end{document}